# Mitigating laser imprint with a foam overcoating

D. X. Liu,[1] T. Tao,[1*] J. Li,[1,2] Q. Jia,[1,2] and J. Zheng [1,2†]

[1]   (Department of Plasma Physics and Fusion Engineering and CAS Key Laboratory of Geospace Environment, University of Science and Technology of China, Hefei 230026, China)

[2]   (Collaborative Innovation Center of IFSA, Shanghai Jiao Tong University, Shanghai, 200240, China)

**Abstract**

Foam has been suggested to reduce laser imprint because of its low density. In this paper, the two-dimensional radiation hydrodynamic code FLASH is applied to investigate and characterize the strength of laser imprint through analyzing areal density perturbation. There are two important factors for the mitigation of laser imprint besides the thermal smoothing of the conduction region (between the ablation front and the critical density surface) and the mass ablation of the ablation front. First, radiation ablation dynamically modulates density distribution not only to increase the frequency of the perturbed ablation front oscillation but also to decrease the amplitude of oscillation. Second, a larger length of the shocked compression region reduces the amplitude of the perturbed shock front oscillation. The smaller the perturbation of both ablation front and shock front, the smaller the areal density perturbation. Based on the above physical mechanisms, the optimal way of mitigating laser imprint with foam is that the dynamically modulated density distribution further reduces the amplitude of perturbation reaching the solid CH when the areal density perturbation of foam oscillates to the first minimum value. The optimal ranges of foam parameters to mitigate laser imprint are proposed with the aid of dimensional analysis: the foam thickness is about 2~3 times the perturbation wavelength, and the foam density is about 1/2~3/2 times the mass density corresponding to the critical density.

[*] tt397396@ustc.edu.cn

[†] jzheng@ustc.edu.cn

## I. Introduction

There are mainly two approaches to inertial confinement fusion (ICF): indirect drive[2] and direct drive[3]. In direct laser-driven ICF, multiple laser beams are irradiated onto a target directly without converting its energy to X-rays in the hohlraum like indirect drive, and the energy coupling efficiency between the laser and the target is higher. However, the target usually suffers a worse uniformity of laser illumination because of beam intensity profile, finite laser beam numbers, focusing geometry[4] and so on. Non-uniformity induces perturbations in the critical density surface as well as in the ablation front, from which a shock propagates into the target resulting in areal density perturbation, of which the process is known as laser imprint. The perturbation induced in the process of laser imprint can become the seed of Rayleigh-Taylor instability (RTI) that occurs during the implosion acceleration phase. If not controlled, it can severely degrade the implosion performance of a target.

In numerical simulations[5,6] and experiments[7,8,9], it has been observed that the areal density perturbation of a target first grows to saturation and then evolves as damped oscillations, which indicates that there are negative feedback stabilization mechanisms in the process of laser imprint. One is due to ablation pressure. When the local distance between the ablation front and the critical density surface is smaller, the heat flow is larger, and the local ablation pressure is higher[10]. Then the local distance intends to increase due to the higher ablation pressure and vice versa. This mechanism causes the perturbed ablation front to oscillate. The other is due to the shock stability. The shock front is stable when it propagates in a normal medium, and oscillates if there are perturbations[11]. The oscillation of the perturbed ablation front and shock front leads to the oscillation of areal density perturbation.

From the above physical mechanisms, it can be seen that areal density perturbation is affected by electron thermal transport and mass ablation. S. E. Bonder[12] proposed that when laser deposition energy is transported from the critical density surface to the ablation front, the lateral thermal diffusion will smooth ablation pressure perturbation. Areal density perturbation that has been imprinted on the target can be alleviated by mass ablation, which is also the main stabilization mechanism of RTI[13,14]. In addition, Freeman's theory[15] points out that the amplitude of shock front perturbation attenuates along with the shock propagation.

The OMEGA facility in the United States is the main platform for direct drive research. Although various advanced techniques are applied for OMEGA to improve beam uniformity, such as phase plate (PP)[16], polarization smoothing (PS)[17], smoothing by spectral dispersion (SSD)[18,19], laser imprint may still lead to shell decompression, early formation of hot spot, mixing of hot spot fuel and cold shell, an increase in hot spot size and other negative effects, and thus degrade neutron yield[20,21]. Therefore, researchers made full use of mitigation mechanisms in the laser imprint and proposed a series of strategies for laser pulses and fusion targets, such as X-ray hybrid direct drive [22,23], medium or high Z doping of target shell[24,25], pre-pulse[26] or shaping pulse[27] before the main pulse, overcoating with a foam shell[28,29,30] and so on. Recently, the novel double cone ignition scheme, which is essentially a kind of direct drive, is proposed by J. Zhang et al[31]. Despite the usage of continuous phase plate (CPP)[16], there are still short-wavelength perturbations as well as long-wavelength perturbations due to the limited number of laser beams and energy imbalance between beams. This means that additional methods of smoothing are required.

In this paper, the two-dimensional Eulerian radiation hydrodynamic code FLASH[1] is applied as a simulation tool. Under the guidance of physical mechanisms of mitigating laser imprint, the electron thermal transport, ablation properties and radiation field effects of CH foam under the action of laser irradiation are studied. With the aid of dimensional analysis, we find the range of CH foam parameters to mitigate laser imprint.

The paper is arranged as follows. In Section II, we give the basic parameter settings of simulations. Simulation results and analysis are presented in Section III. We derive the conditions for the foam thickness and density with the aid of dimensional analysis in Section IV. At last, we draw the conclusion in Section V.

## II. Simulation settings

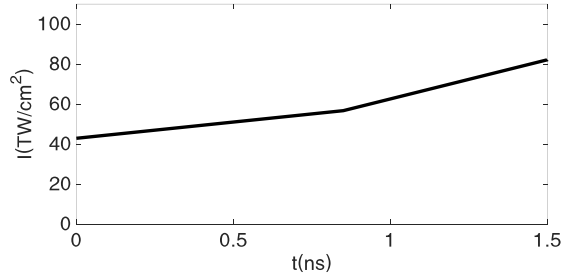

Fig 1. The first 1.5-ns part of the laser pulse in the simulation.

Our simulations are performed in two-dimensional cartesian geometry. A laser beam with a pulse shape indicated in Fig.1 is incident along the negative y-axis with a single-mode cosine ripple in the laser focal spot. The relative perturbation of the laser intensity is $\frac{\delta I_p}{I_p}$=33.5% (ratio of standard deviation to mean), and the perturbation wavelength $\lambda_s$ is 20μm. The laser module adopts a geometric ray tracing algorithm with the only energy deposition mechanism of inverse Bremsstrahlung. We are interested in areal density perturbation which can be measured in experiments. In the post-processing of simulations, the areal density perturbation is defined as, $\delta m(t) = \delta[\sum_{y_3}^{y_2} \rho(x,y,t)]$, $y_2 > y_3$, where ρ is the unit volume mass density, δ[] represents the standard deviation of areal density in the x orientation, $y_3$ is the 70th grid from the ablation front along the negative y-axis, $y_2$ is the 10th grid from the ablation front along the positive y-axis. In this way, all density perturbations in the shocked compression region can be counted. The grid resolutions in the x and y directions are both 0.52μm. The foam is made of CH with a thickness of 60μm and density of 0.025g/cc, and the solid is also CH with a thickness of 45μm and density of 1g/cc.

## III. Simulations and analysis

### A. Simulation results

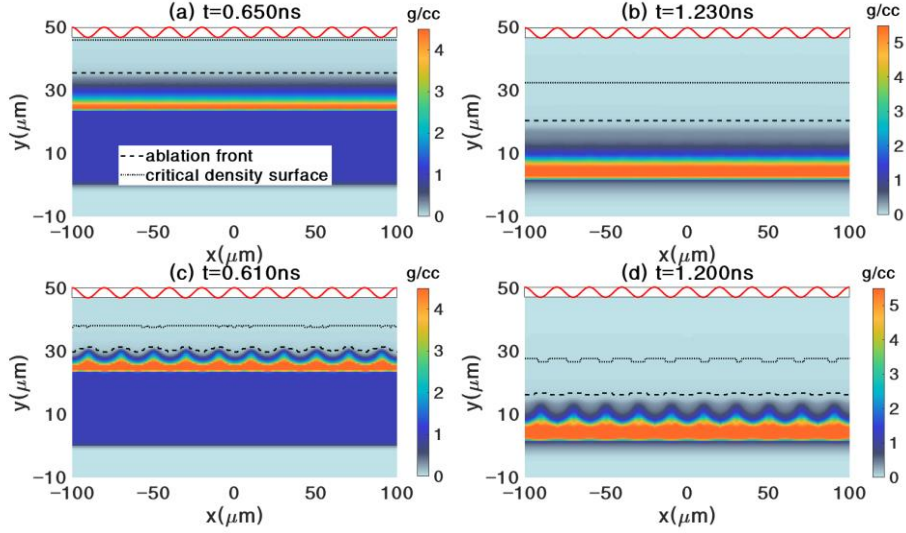

Fig. 2. Density distribution (a) foam-solid CH target, shock front position 22.5μm; (b) foam-solid CH target, shock front position 0μm; (c) bare solid target, shock front position 22.5μm; (d) bare solid target, shock front position 0μm. The inset is normalized laser spatial intensity distribution, $\frac{\delta I_p}{I_p}$=33.5% (ratio of standard deviation to mean). These simulations show that foam is effective in mitigating laser imprint.

First, we show the simulation results of two solid CH targets covered with or without foam. In order to have an intuitive impression of laser imprint, we draw the density distribution when the shock reaches y=22.5μm (1/2 of the thickness of the solid CH) and y=0μm (the rear of the solid CH). In Fig 2, at y=22.5μm, the ablation front of the bare solid target suffers obvious perturbation, which is the dominant factor of areal density perturbation. In contrast, the foam-solid target suffers negligible density perturbation. At y=0μm, the ablation front of the bare solid target becomes smoother, but the density perturbation in the shocked compression region is dominant at this moment. The density perturbation in the foam-solid target is hardly observable. The foam material does take effect in mitigating laser imprint.

The physical process of mitigating laser imprint is manifested with the evolution of ablated plasma. In the foam, the ablation speed ($\sim 10^7 cm/s$) is greater than the local speed of sound ($\sim 10^6 cm/s$), which means that the ablation is supersonic. In Fig 3, the electron thermal ablation front defined at the position of the peak gradient of electron temperature is closer to the solid CH than the shock front defined at the position of peak density. The ablation front perturbation first increases and then decays. This is due to the thermal smoothing and mass ablation on the one hand and the lateral mass perturbation of the compression front (usually the shock front) on the other hand. The saturation time of ablation front perturbation is about 0.05ns, which is much smaller than that of a bare solid target. Therefore, CH foam can quickly respond to laser intensity perturbation, benefiting from higher thermal conductivity and ablation speed, and then the ablation front perturbation rapidly decreases.

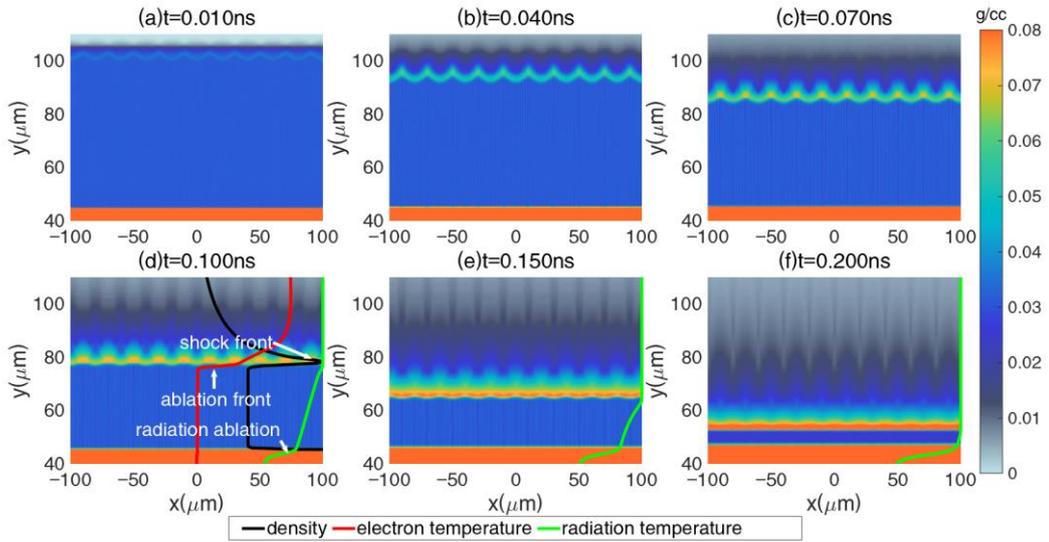

Fig. 3. Foam density evolution with density, electron temperature and radiation temperature line-outs in plot with the scale on the positive x-axis. It shows that foam can quickly respond to laser intensity perturbation, benefiting from the high thermal conductivity and high ablation speed, and then ablation front perturbation rapidly decreases.

The interface between the foam and the solid CH is a contact discontinuity, and has a profound influence on the front of the shock and ablation and the mitigation of laser imprint. Fig. 3 shows that as the foam density is low, there exists radiation ablation occurring at the position of the peak gradient of radiation temperature on the solid CH surface. In Fig. 4, we can see that at 0.23ns, impedance mismatch[34] occurs in the transition region between the foam and the solid CH, causing pressure surge, generation of a reflected shock and an attenuated incident shock. The reflected shock reaches the ablation front and generates a rarefaction wave which propagates in the incident direction of the laser beam. Moreover, when the electron thermal ablation front reaches the radiation ablation region, supersonic ablation gradually evolves into subsonic ablation, and the electron thermal conduction generating shock front is re-formed before the electron thermal ablation front. At 0.27ns, when the electron thermal conduction generating shock front coincides with the radiation shock front, the pressure reaches the maximum, and then gradually decreases.

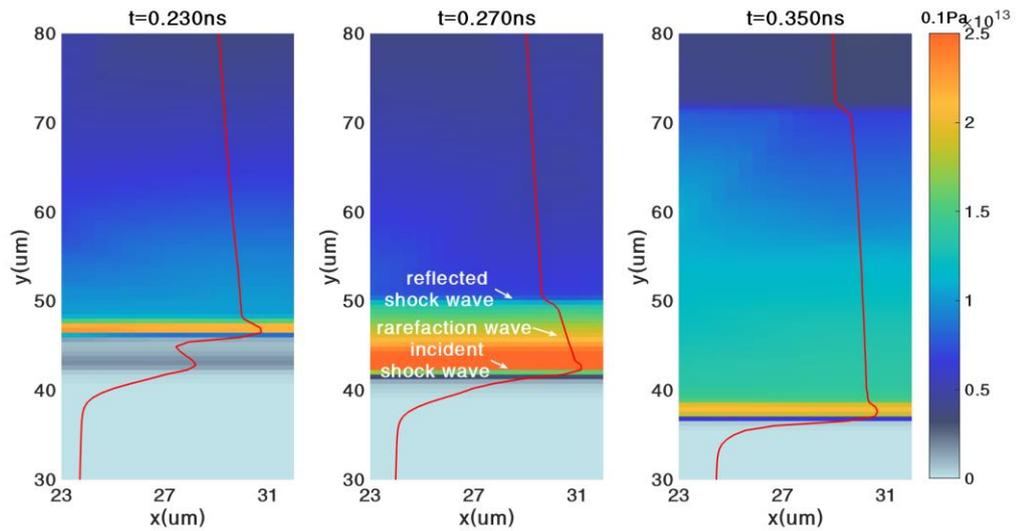

Fig. 4. Pressure distribution in the collision region between the foam and the solid CH with a red pressure line-out like Fig. 3(d). At t=0.23ns, impedance mismatch occurs, causing pressure surge, generation of a reflected shock, an incident shock, and then a rarefaction wave. At t=0.27ns the pressure reaches the maximum and then decrease.

In order to further analyze the mechanisms of mitigating laser imprint, we show the time dependence of the position of the ablation front and the thickness of the thermal conduction zone that is $D_{ac}$. Fig. 5 shows the high thermal conductivity and high ablation rate of foam in detail. As seen in Fig. 5(a), the ablation front locates inside the foam before 0.21ns, and propagates into the solid CH after 0.4ns. From the slope of the trajectory, we can see that the average ablation speed in the foam is large, which is consistent with the equation $v_a = \frac{\dot{m}}{\rho_a}$, where $v_a$ is ablation velocity, $\dot{m}$ is mass ablation rate, and $\rho_a$ is the density of the ablation front. With or without foam, the slope becomes consistent after 0.4 ns, indicating that the foam overcoating does not change the average $v_a$ of the solid CH. Figure 5(b) also shows that the growth rate of $D_{ac}$ in the foam is larger. In t=0.21~0.27ns, $v_a$ decreases rapidly, resulting in a steep decrease in $D_{ac}$. In t=0.27~0.38ns, the reflected shock propagates towards the coronal region with the critical density surface moving along the positive y-axis, and $D_{ac}$ increases. In t=0.38~0.5ns, when $v_a$ is less than that of the critical density surface which also moves along the negative y-axis, $D_{ac}$ decreases. At t=0.1ns, when $D_{ac}$ satisfies the condition $kD_{ac} = 2$, where k is the perturbation wave number, the ablation pressure perturbation induced by the laser intensity perturbation is smoothed in the thermal

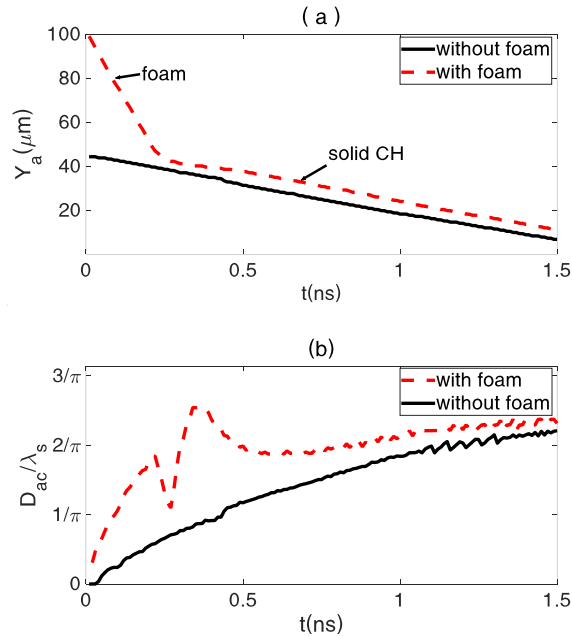

Fig. 5. Time dependence of (a) the position of ablation front with or without foam; (b) the thickness of thermal conduction zone with or without foam. (a) and (b) show the high ablation speed and high thermal conduction of foam.

conduction zone, leaving a negligible influence on areal density perturbation δm, as shown in Fig. 7(a)

Another significant change to the foam-solid target is the thickness of the shocked compression region which is $D_{as}$. In Fig. 6, compared with the bare solid target, the velocity of incident shock increases after the collision, then $D_{as}$ also increases rapidly. As pressure perturbation near the ablation front propagates towards the shock front with local sound speed and decays with time, the larger the $D_{as}$, the smaller the pressure perturbation reaching the shock front, and δm becomes smaller. Finally, with the attenuation of the incident shock, the growth rate of shock compression region is the same as that of the bare solid target.

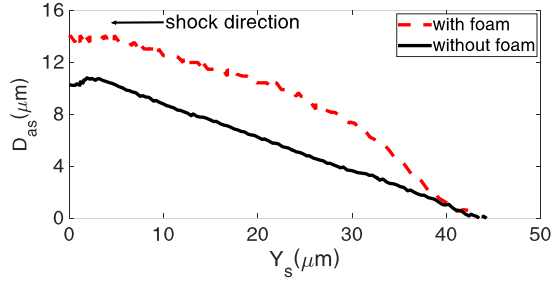

Fig. 6. The thickness of the shocked compression region depending on the shock front position with or without foam. The foam makes the shocked compression region increase and the areal density perturbation decrease.

## B. Discussion and analysis

When the shock passes through the target, there are three sources of δm: the ablation front perturbation, the shock front perturbation, and the acoustic perturbation in the shocked compression region[10]. The expression for the total δm is as follows.

$$\delta m(x,y,t) = \delta \left( \int_{y_1}^{y_a+\eta_a} \rho(x,y,t)\,dx \right)$$
$$= \delta \left( \int_{y_s+\eta_s}^{y_a+\eta_a} \rho_2(x,y,t)\,dx \right) + \delta \left( \int_{y_1}^{y_s+\eta_s} \rho_1(x,y,t)\,dx \right) + \delta \left( \int_{y_s}^{y_a} \tilde{\rho}(x,y,t)\,dx \right) \quad (1)$$

Where $y_1$ is the rear position of the target, $y_a$ and $y_s$ are the average positions of the ablation front and shock front, $\eta_a$ and $\eta_s$ are the perturbations of the ablation front and shock front, respectively, $\rho_1$ and $\rho_2$ are the average densities of the unshocked and shocked area, and $\tilde{\rho}$ is the perturbation of the shocked compression region.

In Fig. 7(a), due to the strong thermal smoothing and mass ablation of the foam, $\eta_a$

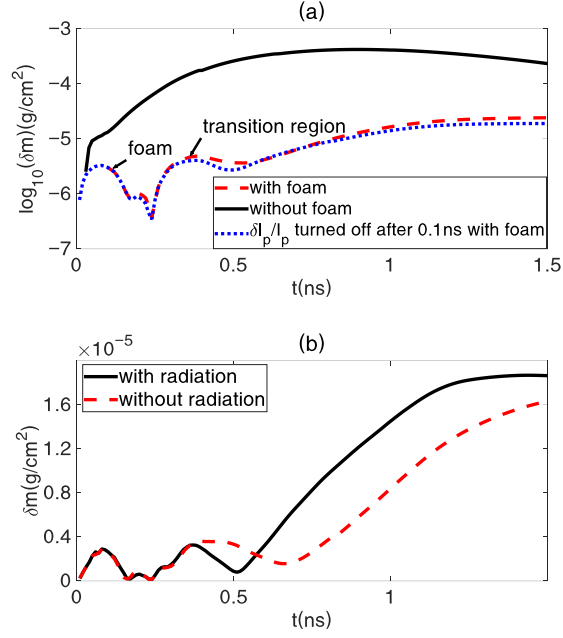

Fig. 7. Time dependence of areal density perturbation (a) with or without foam and the laser intensity perturbation turned off after 0.1ns with foam (logarithmic coordinate); (b) In t=0~0.25ns with or without radiation. For foam with $\rho$=0.025g/cc, thermal conduction is the dominate factor to mitigate laser imprint.

decreases, and the first areal density perturbation oscillation occurs in the foam. The second areal density perturbation oscillation occurs in the transition region where radiation ablation exists. When the length of density distribution in the transition region increases, the ablation front decelerates, and the oscillation frequency of $\eta_a$ and $\delta m$ increases. The oscillation evolution equation of $\eta_a$ is given by[33],

$$\frac{d^2}{dt^2}\eta_a + \frac{4kv_a}{1+r_D}\frac{d}{dt}\eta_a + \left(\frac{d}{dt}kv_a + \frac{k^2v_a^2}{r_D} - Agk\right)\eta_a = 0 \quad (2)$$

Where $r_D = \frac{\rho_{bl}}{\rho_1} < 1$, $\rho_{bl}$ is the density of plasma between the ablation front and critical density surface, $A = \frac{1-r_D}{1+r_D}$ is the Atwood number, g is the effective acceleration of the ablation front, g>0 means that the ablation front decelerates; g=0 means that the speed of the ablation front remains constant. If g≥0, then

$$\left(\frac{4kv_a}{1+r_D}\right)^2 - 4\left(\frac{d}{dt}kv_a + \frac{k^2v_a^2}{r_D} - Agk\right) = -4\left(A^2\frac{k^2v_a^2}{r_D} + (1-A)gk\right) < 0$$

The $\eta_a$ oscillates with the oscillation frequency w and amplitude Am given by.

$$w = \sqrt{A^2\frac{k^2v_a^2}{r_D} + (1-A)gk} \quad ; \quad Am = Ce^{-\frac{2kv_a}{1+r_D}t} \quad (3)$$

Here C is the initial value of $\eta_a$. When g=0, the larger the $v_a$, the greater the oscillation frequency of $\eta_a$; If the initial $v_a$ is the same, the oscillation frequency of $\eta_a$ with g>0 is larger than that of g=0, and the higher frequency oscillation reaches the smaller peak perturbation. That is, $\eta_a$ decreases further during the second areal density perturbation oscillation.

Fig. 7(b) shows that, in t=0~0.25ns, the time dependence of δm is dominated by thermal conduction, while radiation hardly works. After 0.25ns, δm with or without radiation effect is not identical. When the radiation field is artificially turned off in the simulation, density will jump at the interface between the solid CH and the foam, a scale length of density distribution will decrease, oscillation period and perturbation amplitude of δm in the transition region will increase. It can be seen that foam-solid targets with different thicknesses and densities have different radiation ablation and density distribution scale lengths, and get different effects of laser imprint mitigation. At t=1.2ns, the shock breaks through the rear of the target, and δm is not much different, indicating that thermal conduction dominates for the foam overcoating with $\rho$=0.025g/cc. In Fig. 7(a), after 0.5 ns, when $v_a$ in the solid CH is smaller than that of the foam, the oscillation of δm becomes slower, and δm increases. At t=1.2ns, δm remains constant, indicating that the solid CH decompresses under the effect of the rarefaction wave.

In a brief summary of this section, three factors for mitigating laser imprint with foam are identified with the aid of radiation hydrodynamic simulations. First, foam materials have a larger growth rate of thermal conduction zone, larger ablation velocity and smaller ablation front perturbation. Second, the radiation ablation of the solid CH interface dynamically adjusts the scale length of density distribution, making oscillation period and amplitude decrease. Thirdly, when the length of the shocked compression region increases, pressure perturbation from the ablation front to the high-density region near the shock front decreases, and δm also decreases.

**IV. Optimal foam parameters for the mitigation of laser imprint**

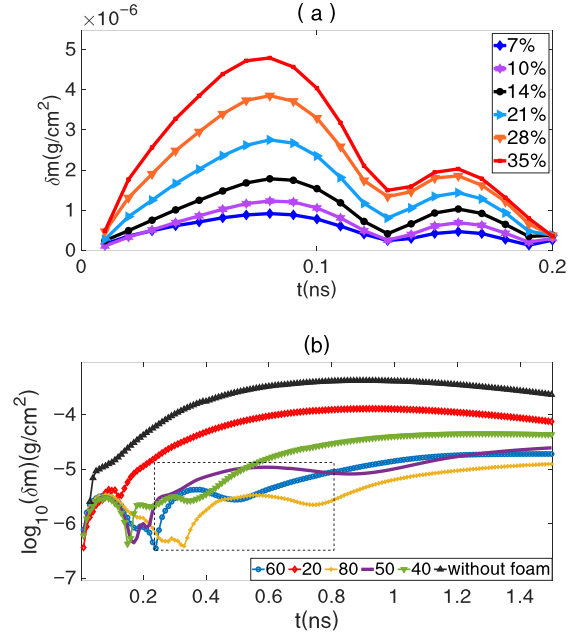

Fig. 8. Time dependence of the areal density perturbation (a) with various laser intensity perturbation, $t_{smooth} \approx 2t_{sat}$, (b) with different foam thickness. The selected region corresponds to areal density perturbation oscillation in the transition region between the solid CH and the foam, which reduces areal density perturbation from the foam to the solid CH.

In Fig. 8(a), δm of the foam induced by laser imprint first increases to saturation whose time is $t_{sat}$ and then attenuates and oscillates. The time when δm reaches the first local minimum is defined as the smoothing time $t_{smooth}$. δm from saturation value to the first local minimum value (process 2) can be regarded as the inverse process from zero value to saturation value (process 1). The average $v_a$ of the two processes is the same, and the thermal smoothing effect of the conduction zone of process 2 is stronger, so $t_{smooth}$ is slightly less than $2t_{sat}$, that is $t_{smooth} \approx 2t_{sat}$. In Fig. 8 (b), the selected region in the frame corresponds to areal density perturbation oscillation in the transition region, which reduces δm from the foam to the solid CH. When the thickness of foam is above 40um, δm is small. The excessive thickness of the foam will reduce energy coupling efficiency between the laser and the solid CH, so the optimal thickness of the foam should be in the range of 40-60um. For the foam with lower density, such as 0.025g/cc, the electron thermal conduction plays a dominant role in mitigating laser imprint. The foam can mitigate perturbation wavelengths of less and greater than 20um.

Regards supersonic ablation, the relationship between the density of the ablation front and foam is $\rho_0 \leq \rho_a \leq 2\rho_0$ [35] and $v_a = \left(\frac{p_a}{\rho_a}\right)^{\frac{1}{2}} \geq \frac{1}{\sqrt{2}}\left(\frac{p_a}{\rho_0}\right)^{\frac{1}{2}}$, where $p_a$ is the ablation pressure. According to dimensional analysis, $t_{sat}$ is proportional to $\lambda_s/v_a$. With the scaling law of unsteady ablation pressure, $p_a \propto I_p^{\frac{3}{4}}\lambda_L^{-\frac{1}{4}}$ [35], where $I_p$ is the average laser power, $\lambda_L$ is the incident laser wavelength, we can get the scaling law of $t_{sat}$,

$$t_{sat} = \frac{\lambda_s}{v_a}T(\mu,\tau) \leq \sqrt{2}\lambda_s\left(\frac{\rho_0}{p_a}\right)^{\frac{1}{2}}T(\mu,\tau) \propto I_p^{-3/8}\lambda_s\rho_0^{1/2}\lambda_L^{1/8}T(\mu,\tau) \qquad (4)$$

Where $\tau = \frac{t_1 v_a}{\lambda}$ and $\mu = \frac{v_a}{v_s}$ are two dimensionless variables, $t_1$ is the thermal smoothing time of conduction zone, $v_s$ is the shock front velocity. $T(\mu,\tau)$ is weakly dependent on the dimensionless variables $\mu$ and $\tau$, and can be approximated as a constant[5].

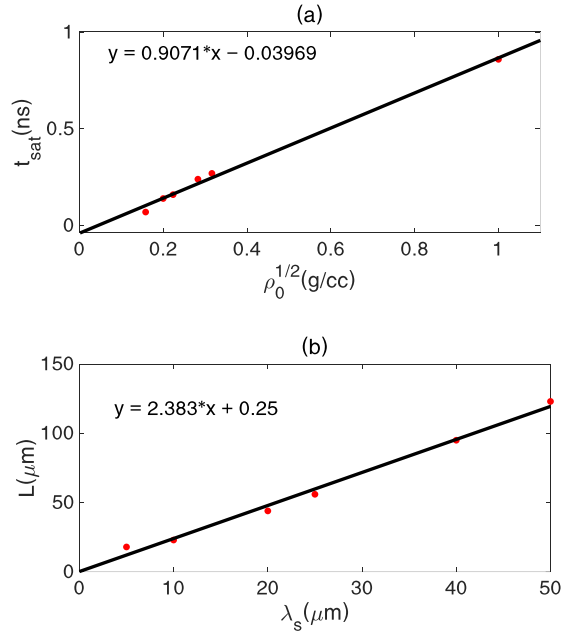

Fig. 9. (a)The saturation time is linearly dependent on the square root of density. (b)The thickness of foam is linearly dependent on the perturbation wavelength.

From the above scaling law (4), we can know that increasing $I_p$ as well as decreasing $\rho_0$, $\lambda_L$ and $\lambda_s$ can reduce $t_{sat}$. In Fig. 9. (a), $t_{sat}$ linearly depends on $\rho_0^{1/2}$, confirming our assumption that $T(\mu,\tau)$ can be approximated as a constant. The thickness of foam required for smoothing density perturbation, which is L, is the product of $t_{smooth}$ and $v_s$. From the scaling raw, $v_s = \sqrt{\frac{\gamma+1}{2}} \cdot \sqrt{\frac{P_a}{\rho_0}}$, $\gamma$ is the ratio of specific heats, L is given by,

$$L = t_{smooth} * v_s \approx 2(1\sim\sqrt{2})\lambda_s * \sqrt{\frac{\gamma+1}{2}} \cdot F(\mu,\tau) = 2(1\sim\sqrt{2})\lambda_s P(\mu,\tau) \quad (5)$$

In Fig. 9(b), L is roughly a linear function of $\lambda_s$ with a slope of 2.4, so $P(\mu,\tau)$ is also weakly dependent on the variables μ and τ, and can be approximated as 1. The optimal L is thus given by $L \approx (2\sim3)\lambda_s$. For example, when $\lambda_s$ is 20um, L should be 40~60um to sufficiently mitigate laser imprint.

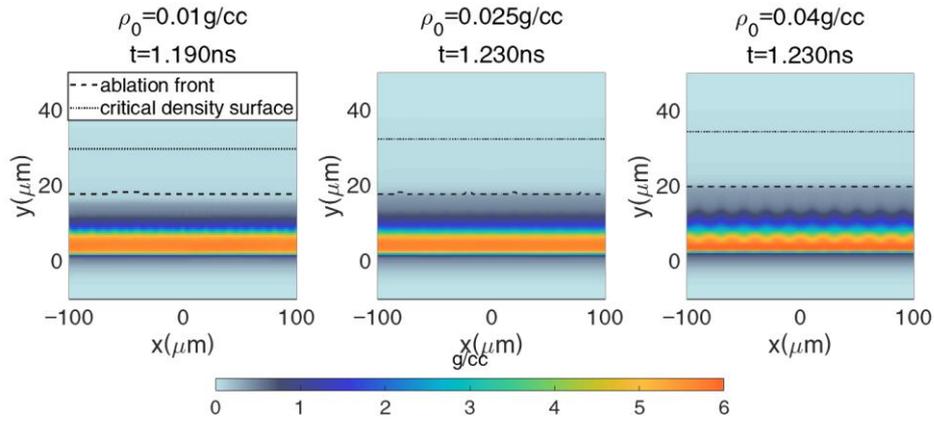

Fig. 10 Density distribution with foam of different densities. With foam of higher density, when thermal conduction effect becomes weaker, radiation effect becomes stronger, and anti-phase density perturbation may occur in the shocked compression region.

The optimal $\rho_0$ should be in the range of $\frac{1}{2}\rho_{cr} < \rho_0 \leq \frac{3}{2}\rho_{cr}$, where $\rho_{cr}$ is the mass density corresponding to the critical density, which depends on the ablated materials. If $\rho_0$ is too low, saying $\rho_0 < \frac{1}{2}\rho_{cr}$, the critical density surface would be located ahead of the ablation front, and thermal smoothing could not play a role to reduce the ablation front perturbation. When $\rho_0$ is too high, saying $\rho_0 > \frac{3}{2}\rho_{cr}$, the time of laser ablating foam with same thickness increases, and the radiation effect of the transition region is enhanced, which causes a long scale length of density and complex oscillation behavior. In this case, thermal conduction competes with radiation effect, which should be balanced to get the optimal. Otherwise anti-phase density perturbation can occur in the shocked compression region, and the amplitude of density perturbation in the solid CH will increases, as shown in Fig. 10. When $\rho_0$ is in the range of $\frac{1}{2}\rho_{cr} < \rho_0 \leq \frac{3}{2}\rho_{cr}$, the perturbation of the critical density surface does not directly contact with solid CH, while thermal smoothing can take into effect due to rapid ablation.

## V. Summary

With the aid of numerical simulations, we investigate in detail the main physical mechanisms of applying foam to mitigate laser imprint. First, when the shock passes through the foam with low density, the thermal smoothing effect of the conduction zone is strong, and the mass ablation speed of the ablation front is large, thus the ablation front perturbation becomes small. Second, during the passing in the transition region between the foam and the solid CH, the radiation ablation dynamically modulates density distribution, increasing the oscillation frequency of the ablation front, and reduces the oscillation amplitude. Thirdly, when the shock passes through the solid CH, the thickness of shocked compression region increases, and the pressure perturbation reaching the shock front reduces which causes smaller shock front perturbation. The smaller the perturbation of both the ablation front and shock front, the smaller the areal density perturbation. Based on the above physical mechanisms, we get the optimal way to mitigate laser imprint with foam. When the oscillation of areal density perturbation in the foam reaches the first minimum value, the oscillation of areal density perturbation in the transition region further reduces perturbation, The ranges of foam parameters are proposed with the aid of dimensional analysis, the thickness of foam is about 2~3 times the perturbation wavelength, and the density is within 1/2~3/2 times the mass density corresponding to the critical density.

## Acknowledgements

This work is supported by the Strategic Priority Research Program of Chinese Academy of Sciences (XDA250100200).

## Reference

1    B. Fryxell, K. Olson, P. Ricker, et al., The Astrophysical Journal Supplement Series 131(1) 273 (2000).
2    J. Lindl, Phys. Plasmas 2 3933 (1995).
3    R. S. Craxton , K. S. Anderson, T. R. Boehly, et al., Phys. Plasmas 22 110501 (2015).
4    S. Skupsky, K. Lee, J. Appl. Phys. 54 3662 (1983).
5    R. Ishizaki, K. Nishihara, Phys. Rev. Lett. 78 1920 (1997).
6    A. L. Velikovich, J. P. Dahlburg, J. H. Gardner, and R. J. Taylor, Phys. Plasmas 5 1491 (1998).
7    Y. Aglitskiy, A. L. Velikovich, M. Karasik, et al., Phys. Rev. Lett. 87 265001 (2001).
8    Y. Aglitskiy, A. L. Velikovich, M. Karasik, et al., Phys. Rev. Lett. 87 265002 (2001).
9    Y. Aglitskiy, N Metzler, M. Karasik, et al., Phys. Plasmas 13 080703 (2006).
10   V. N. Goncharov, O. V. Gotchev, E. Vianello, et al., Phys. Plasmas 13 012702 (2006).
11   L. Landau, E. Lifshitz, Fluid Mechanics (Second Edition). Pergamon, (1987).
12   S. E. Bodner, J. Fusion Energy 1 221 (1981).
13   H. Takabe, L. Montierth, R. L. Morse, Phys. Fluids 26 2299 (1983).

14  H. Takabe, K. Mima, L. Montierth, and R. L. Morse, Phys. Fluids 28 3676 (1985).

15  N. C. Freeman, Proceedings of the Royal Society of London. Series A. Mathematical and Physical Sciences, 228(1174) 341-362 (1955).

16  N. Kang, H. Liu, Y. Zhao, et al., Plasma Phys. Contr. F.62 055007 (2020).

17  T. R. Boehly, V. A. Smalyuk, D. D. Meyerhofer, et al., Journal of applied physics 85(7) 3444-3447 (1999).

18  T. R. Boehly, V. N. Goncharov, O. Gotchev, et al., Phys. Plasmas 8 2331 (2001).

19  S. P. Regan, J. A. Marozas, R. S. Craxton, et al., J. Opt. Soc. Am. B 22(5) 998-1002 (2005).

20  S. X. Hu , D. T. Michel, A. K. Davis, et al., Phys. Plasmas 23 102701 (2016).

21  D. T. Michel, S. X. Hu, A. K. Davis, et al., Physical Review E 95(5): 051202 (2017).

22  J. Oh, A. J. Schmitt, M. Karasik, et al., Phys. Plasmas 28(3) 032704 (2021).

23  M. Karasik, J. Oh, S. P. Obenschain, et al., Phys. Plasmas 28(3) 032710 (2021).

24  G. Fiksel, S. X. Hu, V. A. Goncharov, et al., Phys. Plasmas 19 062704 (2012).

25  S. X. Hu, G. Fiksel, V. N. Goncharov, et al., Physical Review Letters 108(19) 195003 (2012).

26  I. V. Igumenshchev, A. L. Velikovich, V. N. Goncharov, et al., Phys. Rev. Lett 123 065001 (2019).

27  N. Metzler, A. L. Velikovich, A. J. Schmitt, and J. H. Gardner, Phys Plasmas 9 5050 (2002).

28  S. Depierreux, C. Labaune, D. T. Michel, et al., Phys. Rev. Lett 102 195005 (2009).

29  B. Delorme, M. Olazabal-Loumé, A. Casner, et al., Phys. Plasmas 23 042701 (2016).

30  S. X. Hu, W. Theobald, P. B. Radha, et al., Phys. Plasmas 25 082710 (2018).

31  J. Zhang, W. M. Wang, X. H. Yang, et al., Phil.Trans. R. Soc. A 378: 20200015.

32  R. J. Taylor, A. L. Velikovich, J. P. Dahlburg, J. H. Gardner, Phys. Rev. Lett 79(10) 1861 (1997).

33  N. Metzler, A. L. Velikovich, J. H. Gardner, Phys. Plasmas 6 3283 (1999).

34  Y. B. Zel'dovich, Y. P. Raizer, (translated by S. C. Zhang ) Physics of shock waves and high temperature hydrodynamic phenomena (Beijing: Science Press) (1980).

35  S. Atzeni, J. Meyer-Ter-Vehn, (translated by B. F. Shen) The Physics of Inertial Fusion(Beijing: Science Press)p174-177,180-184 (in Chinese)(2008).